\documentclass[prd, aps, superscriptaddress, preprintnumbers, twocolumn, floatfix, nofootinbib]{revtex4}

\usepackage{amsfonts}
\usepackage{amsmath}
\usepackage{amssymb}
\usepackage{bm}
\usepackage{dcolumn}
\usepackage{graphicx}
\usepackage[latin1]{inputenc}
\usepackage{latexsym}
\usepackage{rotating}
\usepackage{hyperref}
\usepackage{graphicx}
\usepackage{color}

\usepackage{amsfonts}
\usepackage{amsmath}
\usepackage{amssymb}
\usepackage{bm}
\usepackage{dcolumn}
\usepackage{graphicx}
\usepackage[latin1]{inputenc}
\usepackage{latexsym}
\usepackage{rotating}
\usepackage{hyperref}

\newcommand\be{\begin{equation}}
\newcommand\ba{\begin{eqnarray}}
\newcommand\ee{\end{equation}}
\newcommand\ea{\end{eqnarray}}

\begin{document}

\title{Eternal Inflation, Entropy Bounds and the Swampland}

\author{Ziwei Wang}
\email{ziwei.wangr@mail.mcgill.ca}
\affiliation{Physics Department, McGill University, Montreal, QC, H3A 2T8, Canada}

\author{Robert Brandenberger}
\email{rhb@physics.mcgill.ca}
\affiliation{Physics Department, McGill University, Montreal, QC, H3A 2T8, Canada,
 and Institute for Theoretical Physics,
ETH Zurich, Wolfgang-Pauli-Strasse 27, 8093, Zurich, Switzerland}

\author{Lavinia Heisenberg} \email{lavinia.heisenberg@phys.ethz.ch}
\affiliation{Institute for Theoretical Physics,
ETH Zurich, Wolfgang-Pauli-Strasse 27, 8093, Zurich, Switzerland}

\date{\today}

\begin{abstract}
 
 
 It has been suggested that low energy effective field theories should satisfy given conditions in order to be successfully embedded into string theory. In the case of a single canonically normalized scalar field this translates into conditions on its potential and the derivatives thereof. In this Letter we revisit stochastic models of small field inflation and study the compatibility of the swampland constraints with entropy considerations. We show that stochastic inflation either violates entropy bounds or the swampland criterium on the slope of the scalar field potential. Furthermore, we illustrate that such models are faced with a graceful exit problem: any patch of space which exits the region of eternal inflation is either not large enough to explain the isotropy of the cosmic microwave background, or has a spectrum of fluctuations with an unacceptably large red tilt.
 
\end{abstract}

\maketitle


\section{Introduction}

The inflationary scenario \cite{Starob,Brout,Guth,Sato} is the current paradigm of early universe cosmology. In addition to explaining the homogeneity, spatial flatness and large size of our universe, the accelerated expansion of space provided by inflation yields a mechanism to explain the origin of structure in the universe \cite{Mukh}. However, inflation is not the only scenario of early universe cosmology which is consistent with current cosmological observations. Alternatives include a bouncing cosmology with a matter-dominated phase of contraction \cite{Fabio}, models based on Born-Infeld inspired modifications of gravity \cite{BeltranJimenez:2017doy}, the Ekpyrotic scenario \cite{Ekp} or an emergent cosmology with initial thermal fluctuations with holographic scaling \cite{NBV}, such as in {\it String Gas Cosmology} \cite{BV} (see e.g. \cite{RHBrev2} for a review of alternatives to cosmological inflation and \cite{Heisenberg:2018vsk} for alternatives of gravity theories relevant for early universe cosmology). Assuming that superstring theory is the fundamental theory which unifies all forces of Nature at a quantum level, it is interesting to ask which (if any) of the currently discussed early universe scenarios emerges from string theory. Observations indicate that our universe is entering another stage of accelerated expansion, the so-called {\it Dark Energy} phase. Another interesting question is how string theory might explain this phase.

Over the past thirty years there has been a lot of work attempting to realize inflation in the context of string theory (see e.g. \cite{Baumann} for an in-depth review). Assuming that space-time is described by General Relativity, scalar field matter is usually used in order to obtain accelerated expansion of space. However,  If superstring theory yields the correct ultraviolet completion of physics at high energy scales, then there are constraints on any effective scalar field model emerging as a low energy description of physics. The criteria on an effective field theory consistent with string theory are called the {\it swampland} criteria (see e.g. \cite{Vafa,Palti} for reviews). Models which do not obey these conditions are said to be in the swampland. It has been shown that these criteria severely constrain inflationary models \cite{swamp3} (see also \cite{other}). Quintessence models \cite{quintessence} of Dark Energy are, at the moment, still viable \cite{swamp3,Lavinia,Linde,Raveri} but will also be severely constrainable using upcoming observations \cite{Lavinia} (see also \cite{Heisenberg:2019qxz}).

The constraints on inflation provided by the swampland criteria have been in general obtained using the classical evolution of scalar fields during inflation. However, quantum fluctuations may have an important effect on scalar field dynamics. According to the {\it stochastic inflation} model \cite{stoch}, quantum fluctuations may counteract the classical force and locally drive the scalar field up the potential, i.e. to larger values of the potential energy density. This is the basis for the {\it eternal inflation} scenario \cite{eternal}. Both in the context of large field inflation \cite{large} and small field inflation \cite{Kinney} it has recently been studied whether these quantum effects can save inflation from the swampland constraints. In the case of large field inflation it was shown that eternal inflation can only be realized if the constant parameter which appears in the swampland constraint for a slowly rolling scalar field (see below) is much smaller than unity, and even in this case only for values of the Hubble expansion rate which are close to the Planck scale, while in the case of small field inflation occuring near a local maximum of the potential a window for eternal inflation consistent with the swampland conjectures was claimed \cite{Kinney}.

In this Letter we study constraints on stochastic models of inflation obtained by combining the swampland constraints with entropy considerations. In analogy to the entropy of a black hole which is given by the area of the event horizon, one can associate an entropy associated with the Hubble horizon $H^{-1}$ (where $H$ is the Hubble expansion rate) of an accelerating universe. In a phase of stochastic inflation the entropy associated with the event horizon decreases in regions where the scalar field moves up the potential. Using a bound on the magnitude of allowed entropy decrease from \cite{Bousso} we show that stochastic inflation either violates this entropy bound, or it violates the swampland criterium on the slope of the scalar field potential. This result reinforces the conclusion that there is tension between the principles of string theory and cosmological inflation.

Attempts to reconcile stochastic inflation with the swampland criteria face another problem, the {\it graceful exit problem}. The density fluctuations which exit the Hubble radius during the period of inflation when stochastic effects dominate are too large in amplitude. Hence, a patch of space in which inflation comes to an end must have undergone a period of slow-roll inflation between when stochastic effects become subdominant and the end of inflation. In the case of large field inflation, the existence of such a phase is inconsistent with the swampland criteria. In the case of small field stochastic inflation \cite{Kinney} we must analyze the problem more carefully. Here we show that for large values of the energy density during the inflationary phase, the rolling phase in islands which exit the eternal inflation region is too short for one Hubble patch exiting the eternal inflation region to become large enough to encompass a universe of our current size. In addition, fluctuations on smaller scales are nonlinear. If the energy density in the inflationary period is lower than a given critical value, a sufficiently long period of evolution after the phase of eternal inflation can be obtained, but the resulting spectrum of fluctuations is far from scale-invariant.

While our manuscript was being completed, two papers appeared which have a large overlap with our work. A first paper \cite{Shandera} also demonstrated that stochastic eternal inflation is in the swampland, focusing on slightly different problems than those we concentrate on. Similar conclusions were reached in \cite{Rudelius} which presented a detailed discussion of the Fokker-Planck equation for stochastic inflation.


\section{Review of the Swampland Criteria}

We will assume that superstring theory is the correct theory of Nature. In this context, scalar fields which arise in the low-energy effective field theory of physics consist of the dilaton, moduli fields and axions. There are many scalar fields which can appear in the low energy effective action, which at first sight appears as good news for scalar field-driven inflation. Howeer, since they all have a particular origin in string theory, their potentials and field ranges in the low energy effective field theory cannot be arbitrary.

The first condition on a scalar field $\phi$ in a low energy effective field theory description of string theory is the field range condition known as {\it distance conjecture} \cite{swamp1} which states that a particular effective field theory has a field range $\Delta \phi$ which is restricted to
\be \label{cond1}
\frac{\Delta \phi}{m_{pl}} \, < \, d \, ,
\ee
where $d$ is a positive constant of the order $1$ and $m_{pl}$ is the four space-time dimensional Planck mass. If we start at a point in field space and move a distance greater than the above one, then new string states will become low mass and have to be included in the low energy effective field theory, thus changing the theory. This condition clearly conflicts with the condition to obtain large-field inflation in canonical scalar field models of inflation since in these models the inflaton field has to move a larger distance in order to obtain a sufficient period of inflation \cite{swamp3}. On the other hand, since Quintessence does not require a large number of e-foldings of accelerated expansion, Quintessence models are not ruled out from the outset \cite{swamp3,Lavinia}.

The second swampland condition \cite{swamp2} applies to situations where a scalar field is rolling while dominating the cosmology. It is a constraint on the slope of the potential of an effective scalar field and states that
\be \label{cond2a}
|\frac{V'}{V}| m_{pl}\, > \, c_1 \, ,
\ee
where $c_1$ is a positive constant of order unity (the prime indicates the derivative with respect to $\phi$). This condition clearly rules out slow roll inflation models with canonical kinetic terms (models with extra friction, e.g. warm inflation \cite{Berera}, can be consistent with this condition \cite{warm}). This condition can be derived \cite{swamp4} by demanding that the entropy obtained by the extra string degrees of freedom which become massless is less than the Gibbons-Hawking entropy \cite{GH} of an accelerating Hubble patch of space. It is applicable provided that the scalar field is in uniform motion, in particular during an epoch of slow roll inflation. However, it is not directly applicable if the scalar field is undergoing stochastic fluctuations without overall slow rolling.

There is a refined version of this swampland condition \cite{swamp4} (see also \cite{other2}) according to which models of effective scalar fields can be consistent with string theory even if the  condition (\ref{cond2a}) is not satisfied or applicable in the region of field space where the dynamics is taking place as long as in this region
\be \label{cond2b}
\frac{V''}{V} m_{pl}^2 \, < - c_2 \, ,
\ee
where $c_2$ is another positive constant of order unity. This condition is applicable if the scalar field starts very close to a local maximum of the potential, or if it undergoes stochastic fluctuations without net rolling. For some applications of this condition to cosmology see e.g. \cite{other3}. Note that these conditions rule out de Sitter solutions, in particular de Sitter solutions for Dark Energy (see also \cite{Dvali} for other arguments for the inconsistency between quantum gravity and de Sitter).

Finally, effective field theories coming from string theory should also obey the {\it weak gravity conjecture} which states that at any point in field space, gravity is the weakest force \cite{WGC}.

The reason why the swampland conditions cannot be seen in the context of pure effective field theory is that at the level of an effective field theory, important string degrees of freedom associated with the string oscillatory and winding modes are not taken into account (see e.g. the detailed discussion in \cite{Palti}). These modes do, indeed, play a crucial role in {\it String Gas Cosmology} \cite{BV}, a model of a stringy early universe based on the new fundamental degrees of freedom and symmetries of string theory which are lost at the level of an effective field theory, and which yields an alternative to the inflationary paradigm of structure formaion \cite{NBV} (see \cite{SGC} for a review and \cite{BNPV} for specific predictions for upcoming observations). A characteristic example of a scalar field in an effective field theory emerging from string theory is a K\"ahler modulus field, which in the simple setup of a toroidal compactification of the extra spatial dimension can be viewed as the radius of an extra cycle. As discussed in \cite{Patil}, the interplay of string winding and oscillatory modes leads to a minimum of the effective potential for this field which is at the string scale. This is an example of how the field distance constraint arises in a particular example. In this same example, the value of the potential at its minimum is zero, and the potential is quadratic about the minimum, thus showing that the criteria (\ref{cond2a}) and (\ref{cond2b}) are satisfied.

At first sight, it appears that in the derivation of the swampland conditions on the scalar field potential $V(\phi)$ it was assumed that $\phi$ obeys the classical equation of motion without any quantum fluctuations. However, it is known that the effective scalar field $\phi$ in any given Hubble patch obtains quantum fluctuations from inhomogeneities of larger wavelength which contribute to the local background. This gives rise to a source term in the effective equation of motion for $\phi$ whose magnitude is given by the Hubble expansion rate $H$ \cite{stoch}. As shown in \cite{large}, inclusion of this stochastic term does not help rescue large-field eternal inflation from the swampland. In \cite{Kinney}, however, it was suggested that the stochastic term may save small-field eternal inflation. This is the claim which we study in the following.

\section{Stochastic Effects}

In the presence of stochastic effects, the equation of motion for the effective homogeneous component of a scalar field in a given Hubble patch is
\be \label{stoch}
{\ddot{\phi}} + 3H {\dot{\phi}} + V' \, = \, {\cal N}(t, x) \, ,
\ee
 and the amplitude of $N$ is determined by having the stochastic term lead
 to a change $\langle\delta \phi^2\rangle = \frac{H^2}{4\pi^2}$ in one Hubble time step
\cite{stoch}.

The classical field variation over one Hubble expansion time $H^{-1}$ is given by
\be \label{classical}
\delta \phi_c \, = \, \frac{{\dot{\phi}}}{H} \, ,
\ee
while the change in $\phi$ induced by the noise over the same time interval is
\be \label{quantum}
\delta \phi_q \, = \, \frac{H}{2\pi} \, .
\ee
The region of eternal inflation holds for field values for which
\be \label{etcond}
|\delta \phi_q| \, > \, |\delta \phi_c| \, .
\ee
Making use of the slow-roll equation of motion for $\phi$ to determine the classical movement of $\phi$ and of the Friedmann equation, coupled with the assumption that the energy density is dominated by the scalar field potential energy in order to solve for $H$ in (\ref{quantum}), the condition (\ref{etcond}) becomes
\be \label{etcond2}
\frac{V'}{V}m_{pl} \, < \, \frac{1}{2\pi} \frac{V^{1/2}}{m_{pl}^2} \, .
\ee
In the range of field values where this condition is satisfied, $\phi$ undergoes stochastic fluctuations without net rolling, and hence the condition (\ref{cond2a}) is not directly applicable. In contrast, (\ref{cond2b}) can be applied.

In the case of a simple quadratic potential
\be
V(\phi) \, = \, \frac{1}{2} m^2 \phi^2 \, ,
\ee
where $m$ is some mass which must be much lower than the Planck mass in order that the induced cosmological fluctuations are compatible with observational bounds, the condition (\ref{etcond2}) becomes
\be
|\phi| \, > \, \sqrt{2\pi} \bigl( \frac{m_{pl}}{m} \bigr)^{1/2} m_{pl} \, .
\ee
For a more general potential for large field inflation of the form
\be
V(\phi) \, = \, m_{pl}^4  f(\phi) \, ,
\ee
where $f(\phi)$ is a dimensionless function, the condition (\ref{etcond2}) reads
\be
\frac{f'}{f}m_{pl} \, < \,  \frac{1}{2\pi} f^{1/2} \, .
\ee

The condition for the scalar field potential obeying the swampland condition (\ref{cond2a}) in the field region where stochastic effects are important becomes
\be
c_1 \, < \frac{V'}{V}m_{pl} \, < \, V^{1/2} m_{pl}^{-2} \, ,
\ee
which for values of $c_1$ of order 1 excludes inflation in field regions where effective field theory can be applied.

\section{Entropy Bounds, Inflation and the Swampland}

Another way to see the incompatibility of large field inflation and the swampland constraints is by considering the Gibbons-Hawking \cite{GH} entropy of the local
event horizon of an accelerating cosmology. This entropy is given by the
area of the Hubble horizon in Planck units \cite{Bousso}, i.e.
\be
S_{GH} \, = \,  4\pi H^{-2} m_{pl}^2 \, .
\ee
During a time interval in a stochastic inflation phase when the scalar field moves up the potential due to quantum fluctuation, $H$ increases and hence the entropy bounded
by the area of the horizon decreases. This process may violate the second law of thermodynamics.
Even taken quantum effect of gravity into consideration, the decreases should not be larger than $\mathcal O(1)$ in Planck units.
\be \label{entcond}
\delta S \, > \, -1 \, .
\ee

Let us consider a patch in which the scalar field is moving up the potential.
In this case, the field jump induced by quantum fluctuations is larger in magnitude than the jump induced by the classical force, and we can write down a lower bound on the magnitude of the field jump by taking the classical jump (\ref{classical}) evaluted using the slow-roll equation. The change in the entropy of the patch in one Hubble interval (the coherence time of the quantum fluctuations) can hence be bounded by
\be
\delta S \, \sim \, \delta\bigl( \frac{1}{H^2} \bigr) m_{pl}^2 \, = \, \delta \bigl( \frac{3 m_{pl}^4}{V}\bigr) \, ,
\ee
where in the last step we have used the Friedmann equation. Taking the variation of the last term yields
\be
\delta S \, \sim \, - \frac{3 m_{pl}^4}{V} \frac{V'}{V} \delta \phi \, .
\ee
Considering the change of the entropy in a coherence time during which $\delta \phi = H/(2\pi)$, the entropy condition (\ref{entcond}) then yields
\be
\frac{V'}{V} m_{pl} \, < \frac{H}{m_{pl}} \, ,
\ee
which again shows that inflation at sub-Planckian densities is inconsistent with the swampland condition (\ref{cond2a}) for values of the constant $c_1$ of the order 1.

\section{Graceful Exit Problem for Small Field Eternal Inflation}

To be specific, let us consider the following potential
\be
V(\phi) \, = \, V_0 \cos(\phi / \mu) \, ,
\ee
where $\mu$ determines the curvature of the potential, and $V_0$ its absolute value. Small field inflation takes place while $\phi$ is close to a local maximum of the potential (e.g. $\phi = 0$). When expanded about that point we get the same potential as was considered in \cite{Kinney}), namely
\be \label{pot}
V(\phi) \, \simeq \, V_0 \bigl( 1 - \frac{1}{2} (\frac{\phi}{\mu})^2 \bigr)
\ee
(for $|\phi| \ll \mu$). The swampland condition (\ref{cond2b}) is satisfied provided that
\be \label{range1}
\mu \, < \, \bigl( \frac{1}{c_2} \bigr)^{1/2} m_{pl} \, .
\ee
The field region where eternal inflation is possible is given by (see (\ref{etcond2}))
\be
|\phi| \, < \, \phi_c \, \equiv \, \frac{3}{2} V_0^{1/2} \mu^2 \, ,
\ee
(in Planck units). For eternal inflation to work, this field range needs to be larger than the
size of the quantum field fluctuation. Making use of (\ref{quantum}) this implies
\be \label{range2}
\mu \, > \, \bigl( \frac{2}{6\pi} \bigr)^{1/2} \, ,
\ee
(again in Planck units). Comparing (\ref{range1}) and (\ref{range2}) we see that for $c_2 < 6\pi$
there is a small region for $\mu$ in which both conditions can be satisfied. In the following we
will assume that we are in this region of $\mu$ space.

In order to obtain a period of slow-roll inflation for $|\phi| > \phi_c$ there is an additional lower bound on $\mu$
\be
\mu \, > \, \frac{\sqrt{2}}{3} \, ,
\ee
which can be derived by solving the slow-roll equation of motion for $\phi$ (in the approximation when $H$ is treated as constant) and checking for self-consistency.

Expressed in tems of the e-folding number $N$, the stochastic equation takes the form
 \be \label{Neq}
 H^2 \phi''(N) + 3 H^2 \phi'(N) - V_0 \frac{\phi}{\mu^2} \, = \, \frac{3 H^3}{2\pi} \xi(N) \, ,
 \ee
where $\xi(N)$ is a Gaussian random variable with mean zero and unit variance, i.e.
\be
<\xi(N)> \, = \, 0 \, , \,\,\,\,\, <\xi(N) \xi(N')> \, = \, \delta(N - N') \, .
\ee

We performed simulations of the time evolution of the scalar field $\phi$ in the presence of the stochastic noise we have described. The parameters chosen were $V_0 = 10^{-8}$, $\mu = \sqrt{2} / 3^{1/4}$ and $\phi(0) = 10^{-3}$  (all in Planck units). In Figure 1, the vertical axis is the probability distribution of $\phi$ after $N = 6$ numbers of e-folding as a function of the field value (horizontal axis). By numerical solving the
differential equation (\ref{Neq}) with a stochastic source term in discrete time steps, we may compute the
distribution of physical observables such as the field value at a certain time.
The graph is based on 1000 simulations. The results are scattered about the value which would be obtained without the stochastic term.
\begin{figure}[h!]
 \includegraphics[width=\hsize]{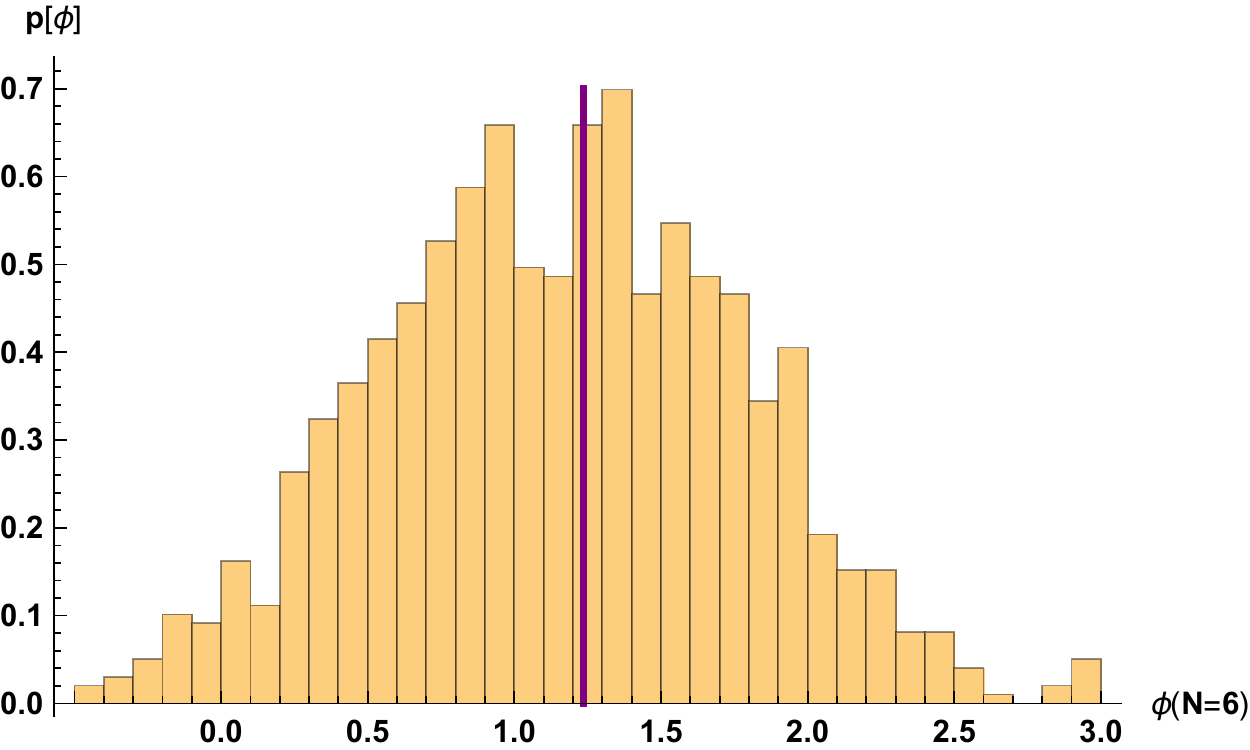}
 \caption{Probability distribution of the field value at e-folding number $6$ based on 1000 simulations with the parameters $V_0 = 10^{-8}, \mu = \frac{\sqrt{2}}{3^{1/4}}, \phi(0) = 10^{-3}$. The value of $H$ was computed assuming that the potential energy dominates. The purple line indicates the value obtained neglecting the stochastic term. }
 \label{fig-field-dist}
\end{figure}
Figure 2 shows the resulting distribution (vertical axis) of the {\it slow-roll parameter} $\epsilon$
for the same simulation set.
\begin{figure}[h!]
 \includegraphics[width=\hsize]{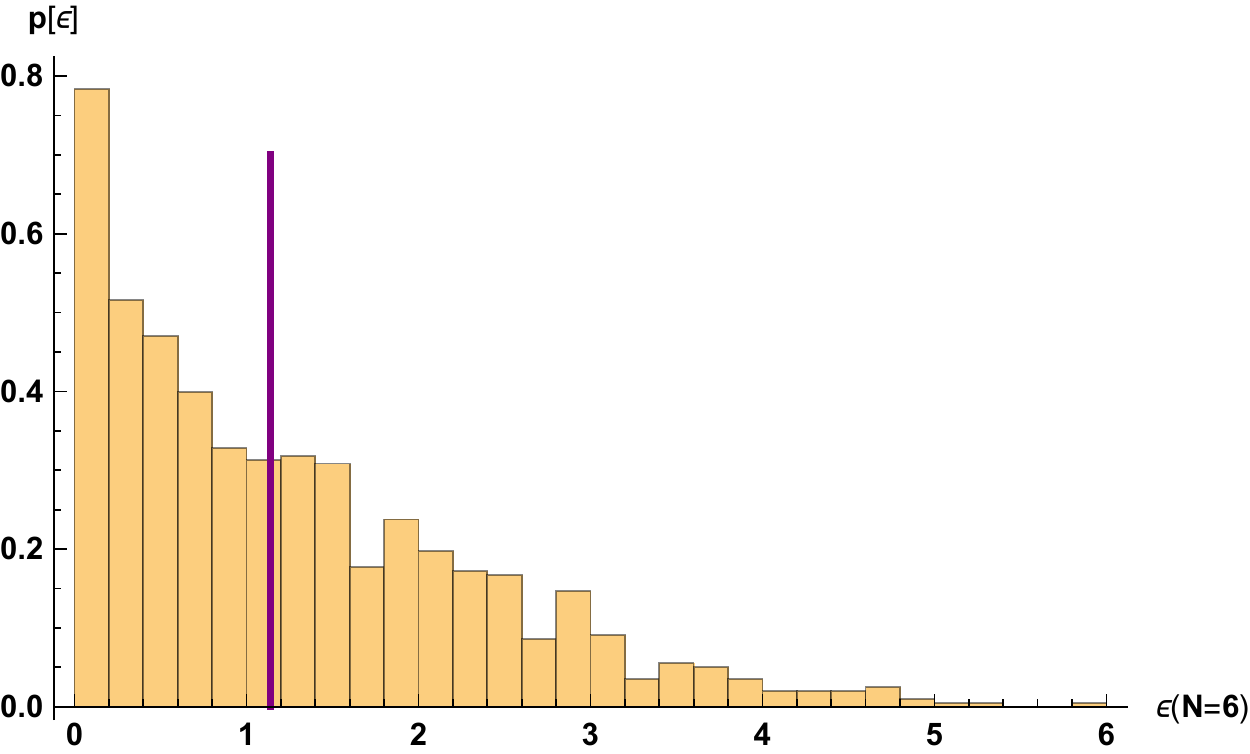}
 \caption{Probability distribution of the first slow-roll parameter for $N = 6$ in the same set of simulatioins as in Figure 1.}
 \label{fig-field-dist}
\end{figure}
In Figure 3 we show the distribution of number of e-foldings until the end of inflation, again for the same simulation set as Figures 1 and 2.
\begin{figure}[h!]
 \includegraphics[width=\hsize]{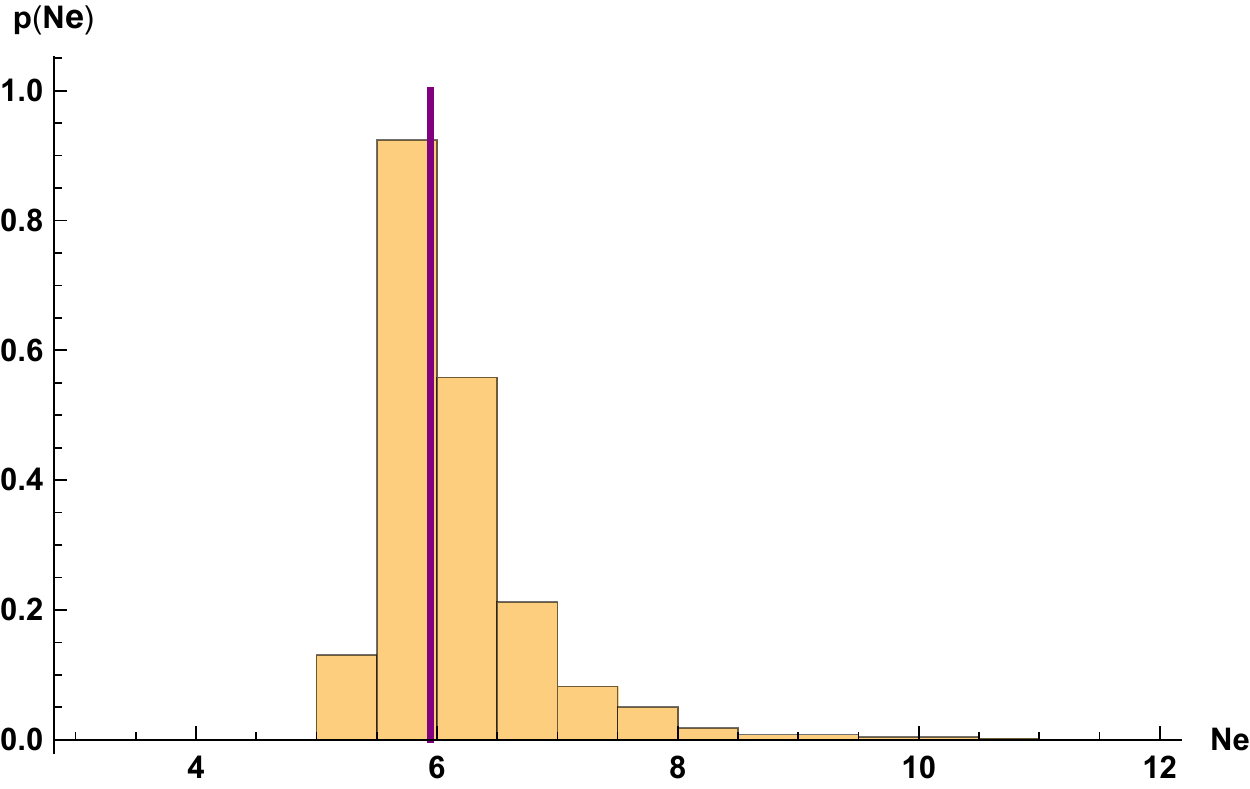}
 \caption{Probablity distribution of the duration of the period of slow-roll inflation in the same set of simulations described in Figure 1.}
 \label{fig-field-dist}
\end{figure}

In the following we will estimate the period of evolution in the post-eternal phase, i.e. for $|\phi| > \phi_c$. In this phase we can neglect the stochastic term. We treat $H$ as constant (thus obtaining an upper bound on the period of evolution since we are taking an upper bound on the friction coefficient). In this approximation, the equation of motion for $\phi$ has exponential solutions
\be
\phi(t) \, \sim e^{\alpha (t - t_c)} \, ,
\ee
where $t_c$ is the initial time in this period (when $|\phi| = \phi_c$), and the two solutions for the index $\alpha$ are
\be
\alpha_{\pm} \, = \, V_0^{1/2} \bigl[\pm \sqrt{\frac{9}{4} + 2 \frac{1}{\mu^2}} - \frac{3}{2} \bigr] \, .
\ee
The dominant solution has index
\be
\alpha_{+} \, = \, \frac{3}{2} V_0^{1/2} \gamma \, ,
\ee
where
\be
\gamma \, = \, \bigl( \sqrt{1 + \frac{8}{9} \frac{1}{\mu^2}} - 1 \bigr) \, .
\ee
To obtain an upper bound on the duration of this phase, we can compute the time $\Delta t = t - t_c$ it takes for the field to evolve from $\phi = \phi_c$ to $\phi = \mu$, assuming that it evolves by the dominant mode. The result is
\be \label{duration}
N \, \equiv \, \Delta t H \, = \, \frac{1}{\gamma} {\rm{ln}} \bigl( \frac{2}{3 \mu V_0^{1/2}} \bigr) \, ,
\ee
(where $H$ is taken to be the initial value of $H$).

Figure 4 shows the time evolution of the scalar field once it exits the region of eternal inflation and stochastic terms can be neglected. The horizontal axis is time (measured in terms of $N(t)$, the vertical axis gives the field value in Planck units. 
The blue (bottommost) curve shows the evolution of $\phi$ including the stochastic noise term, but with the direction of the noise always pointing against that of the classical force. This curve gives an upper bound on the number of e-foldings. The green curve (next one from below) shows the result of an analytical estimate of the evolution of $\phi$ neglecting the noise term, keeping $H$ to be constant, and including both modes of $\phi$ (not only the growing mode as we have above). The result obtained is almost the same as what is obtained by solving the classical equation exactly (taking the time dependence of $H$ into account). This classical evolution is shown in the yellow curve (which deviates from the green one only at the very end).  Finally, the topmost (orange) curve shows our condition for the end of inflation.

\begin{figure}[h!]
 \includegraphics[width=\hsize]{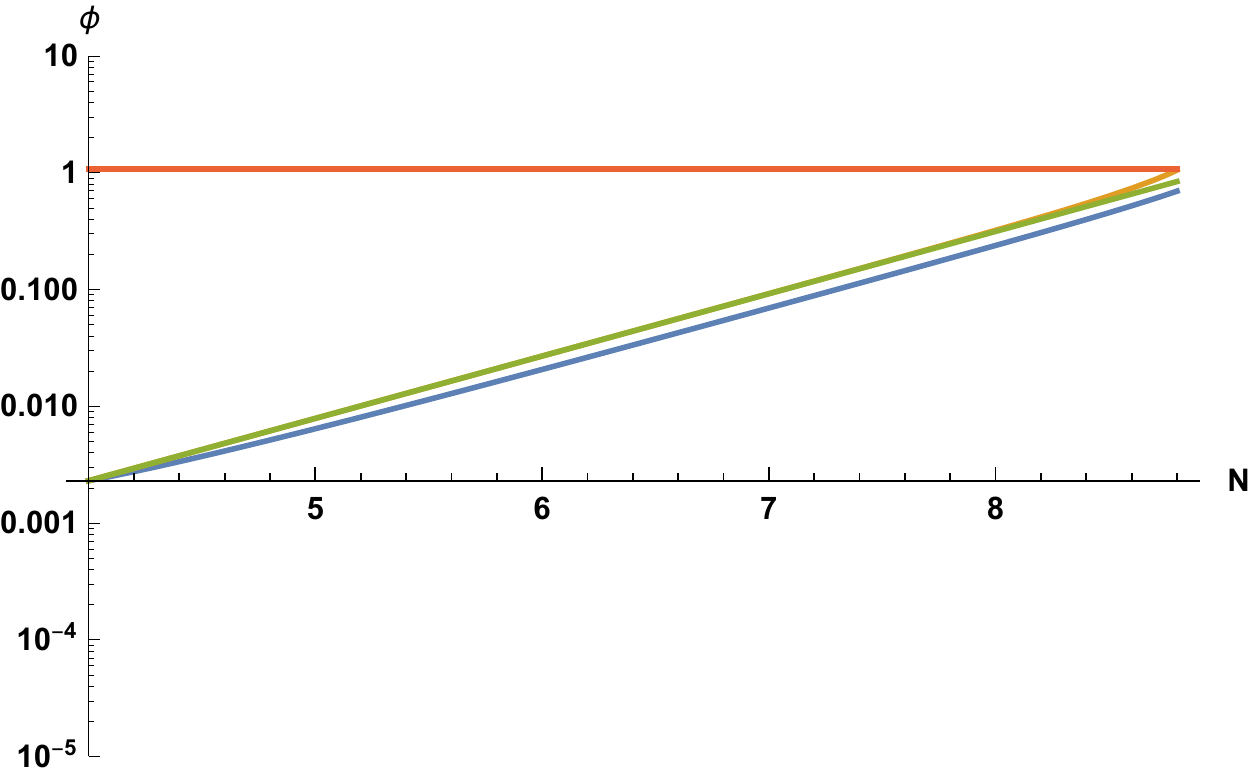}
 \caption{Time evolution of the scalar field once it exits the region of eternal inflation. The horizontal axis is time as measured in terms of e-folding number, the vertical axis gives the field value in Planck units. The approximations which yield the various curves are described in the text.}
 \label{fig-e-folding}
\end{figure}

We see that for values of $V_0$ similar to those which are usually used to obtain inflationary evolution (those close to the scale of Grand Unification), the period is less than $N = 50$. Hence, modes which are observed today in the microwave background and in the large-scale structure crossed the Hubble radius during the period of stochastic inflation. This leads to a serious problem for the amplitude of cosmological fluctuations. The amplitude of the curvature power spectrum is given by (see e.g. \cite{MFB} for a review, and \cite{RHBrev} for a summary discussion)
\be
P(k) \, = \, \bigl( \frac{H^2}{2 \pi {\dot{\phi}}(t_k)} \bigr)^2 \, ,
\ee
where $k$ is the wavenumber and $t(k)$ is the time when the wavelength crosses the Hubble radius. However, during the stochastic phase this quantity is comparable or larger than unity.

For values of $V_0$ small enough such that the value of $N$ from (\ref{duration}) is larger than about $50$, i.e. for
\be \label{small}
V_0^{1/2} \, < \, e^{- 50\gamma} \frac{2}{3 \mu} \, ,
\ee
the amplitude of $P(k)$ can be made sufficiently small. However, the spectrum is not scale-invariant. It is given by
\ba
P(k) \, &=& \, \frac{1}{\mu^2}  (3 \pi \gamma)^{-2} e^{2 \gamma N(k)} \nonumber \\
&=& \,  \frac{1}{\mu^2}  (3 \pi \gamma)^{-2} \bigl( \frac{H}{k} \bigr)^{2\gamma} \, ,
\ea
where $N(k)$ is the number of e-foldings of evolution before the mode $k$ exits the Hubble radius, and we are working in the approximation that $H$ is constant (if we drop this approximation, then the spectrum will be even farther from scale-invariant). Since $\gamma$ is of the order one, the slope of the spectrum is inconsistent with observations.

Note that there is a more serious problem in the above case of small $V_0$: during the post-stochastic period of slow roll inflation the entropy criterion of \cite{swamp4} can be applied, and we conclude that the model lies in the swampland provided that (\ref{cond2a}) is violated, which is the case if $V_0 < 1$ in Planck units.
\\

\section{Conclusions and Discussion}

We have considered the possibility that eternal inflation might be consistent with the swampland conditions. In the case of large field inflation we have shown using several arguments, in particular a quantum bound on the magnitude of allowed entropy decrease in a Hubble patch, that inflation is in conflict with the swampland conjectures. In the case of small field inflation there is a small window in parameter space of typical potentials such as (\ref{pot}) where a phase of eternal accelerated expansion could be made consistent with the swampland constraints. However, such a scenario suffers from the graceful exit problem: typically the size of the patch of the universe which exits the eternal inflation region is too small to be compatible with observations, a similar problem to the one which the original scalar field-driven model of inflation suffered from \cite{Guth2}. This problem can be avoided if the energy scale of the eternal inflation region is sufficiently small, as given by (\ref{small}). However, in this case the induced spectrum of cosmological perturbations is red and inconsistent with observations. Furthermore, the constraint (\ref{cond2a}) can be applied during the phase of slow rolling and the model can then be shown to lie in the swamp for $V_0 < 1$.

\section*{Acknowledgements}

The research at McGill is supported in part by funds from NSERC and from the Canada Research Chair program.
LH is supported by funding from the European Research Council (ERC) under the European Union Horizon 2020 research and innovation programme grant agreement No 801781 and by the Swiss National Science Foundation grant 179740.


\begin{thebibliography}{99}
 
 \bibitem{Starob}
 A.~A.~Starobinsky,
 ``A New Type Of Isotropic Cosmological Models Without Singularity,''
 Phys.\ Lett.\ B {\bf 91}, 99 (1980).
 
 \bibitem{Brout}
 R.~Brout, F.~Englert and E.~Gunzig,
 ``The Creation Of The Universe As A Quantum Phenomenon,''
 Annals Phys.\  {\bf 115}, 78 (1978).
 
 \bibitem{Guth}
 A.~H.~Guth,
 ``The Inflationary Universe: A Possible Solution to the Horizon and Flatness Problems,''
 Phys.\ Rev.\ D {\bf 23}, 347 (1981)
 [Adv.\ Ser.\ Astrophys.\ Cosmol.\  {\bf 3}, 139 (1987)].
 doi:10.1103/PhysRevD.23.347
 
 \bibitem{Sato}
 K.~Sato,
 ``First Order Phase Transition Of A Vacuum And Expansion Of The Universe,''
 Mon.\ Not.\ Roy.\ Astron.\ Soc.\  {\bf 195}, 467 (1981).
 
 \bibitem{Mukh}
 V. Mukhanov and G. Chibisov,
 ``Quantum Fluctuation And Nonsingular Universe. (In Russian),''
 JETP Lett.\  {\bf 33}, 532 (1981) [Pisma Zh.\ Eksp.\ Teor.\ Fiz.\  {\bf 33}, 549 (1981)].
 
 \bibitem{Fabio}
 F.~Finelli and R.~Brandenberger,
 ``On the generation of a scale-invariant spectrum of adiabatic  fluctuations in cosmological
 models with a contracting phase,''
 Phys.\ Rev.\  D {\bf 65}, 103522 (2002) [arXiv:hep-th/0112249].
 
 
 \bibitem{BeltranJimenez:2017doy}
 J.~Beltran Jimenez, L.~Heisenberg, G.~J.~Olmo and D.~Rubiera-Garcia,
 ``Born-Infeld inspired modifications of gravity,''
 Phys.\ Rept.\  {\bf 727}, 1 (2018)
 [arXiv:1704.03351 [gr-qc]].
 
 
 \bibitem{Ekp}
 J.~Khoury, B.~A.~Ovrut, P.~J.~Steinhardt and N.~Turok,
 ``The Ekpyrotic universe: Colliding branes and the origin of the hot big
 bang,''
 Phys.\ Rev.\ D {\bf 64}, 123522 (2001) [hep-th/0103239].
 
 \bibitem{NBV}
 A.~Nayeri, R.~H.~Brandenberger and C.~Vafa,
 ``Producing a scale-invariant spectrum of perturbations in a Hagedorn phase of string
 cosmology,''
 Phys.\ Rev.\ Lett.\  {\bf 97}, 021302 (2006) [arXiv:hep-th/0511140].
 
 \bibitem{BV}
 R.~H.~Brandenberger and C.~Vafa,
 ``Superstrings In The Early Universe,'' Nucl.\ Phys.\ B {\bf 316}, 391 (1989).
 
 \bibitem{RHBrev2}
 R.~H.~Brandenberger,
 ``Alternatives to the inflationary paradigm of structure formation,''
 Int.\ J.\ Mod.\ Phys.\ Conf.\ Ser.\  {\bf 01}, 67 (2011)
 doi:10.1142/S2010194511000109
 [arXiv:0902.4731 [hep-th]].
 
 \bibitem{Heisenberg:2018vsk}
 L.~Heisenberg,
 ``A systematic approach to generalisations of General Relativity and their cosmological implications,''
 Phys.\ Rept.\  {\bf 796}, 1 (2019)
 [arXiv:1807.01725 [gr-qc]].
 
 \bibitem{Baumann}
 D.~Baumann and L.~McAllister,
 ``Inflation and String Theory,''
 doi:10.1017/CBO9781316105733
 arXiv:1404.2601 [hep-th].
 
 \bibitem{Vafa}
 T.~D.~Brennan, F.~Carta and C.~Vafa,
 ``The String Landscape, the Swampland, and the Missing Corner,''
 PoS TASI {\bf 2017}, 015 (2017)
 doi:10.22323/1.305.0015
 [arXiv:1711.00864 [hep-th]].
 
 \bibitem{Palti}
 E.~Palti,
 ``The Swampland: Introduction and Review,''
 arXiv:1903.06239 [hep-th].
 
 \bibitem{swamp3}
 P.~Agrawal, G.~Obied, P.~J.~Steinhardt and C.~Vafa,
 ``On the Cosmological Implications of the String Swampland,''
 Phys.\ Lett.\ B {\bf 784}, 271 (2018)
 doi:10.1016/j.physletb.2018.07.040
 [arXiv:1806.09718 [hep-th]].
 
 \bibitem{other}
 A.~Achucarro and G.~A.~Palma,
 ``The string swampland constraints require multi-field inflation,''
 arXiv:1807.04390 [hep-th];\\
 J.~L.~Lehners,
 ``Small-Field and Scale-Free: Inflation and Ekpyrosis at their Extremes,''
 JCAP {\bf 1811}, no. 11, 001 (2018)
 doi:10.1088/1475-7516/2018/11/001
 [arXiv:1807.05240 [hep-th]];\\
 I.~Ben-Dayan,
 ``Draining the Swampland,''
 arXiv:1808.01615 [hep-th];\\
 W.~H.~Kinney, S.~Vagnozzi and L.~Visinelli,
 ``The Zoo Plot Meets the Swampland: Mutual (In)Consistency of Single-Field Inflation, String Conjectures, and Cosmological Data,''
 arXiv:1808.06424 [astro-ph.CO];\\
 H.~Murayama, M.~Yamazaki and T.~T.~Yanagida,
 ``Do We Live in the Swampland?,''
 JHEP {\bf 1812}, 032 (2018)
 doi:10.1007/JHEP12(2018)032
 [arXiv:1809.00478 [hep-th]];\\
 S.~Brahma and M.~Wali Hossain,
 ``Avoiding the string swampland in single-field inflation: Excited initial states,''
 arXiv:1809.01277 [hep-th];\\
 C.~Damian and O.~Loaiza-Brito,
 ``Two-field axion inflation and the swampland constraint in the flux-scaling scenario,''
 Fortsch.\ Phys.\  {\bf 67}, no. 1-2, 1800072 (2019)
 doi:10.1002/prop.201800072
 [arXiv:1808.03397 [hep-th]];\\
 S.~Das,
 ``A note on Single-field Inflation and the Swampland Criteria,''
 arXiv:1809.03962 [hep-th];\\
 D.~Wang,
 ``The multi-feature universe: large parameter space cosmology and the swampland,''
 arXiv:1809.04854 [astro-ph.CO];\\
 H.~Fukuda, R.~Saito, S.~Shirai and M.~Yamazaki,
 ``Phenomenological Consequences of the Refined Swampland Conjecture,''
 arXiv:1810.06532 [hep-th];\\
 R.~Schimmrigk,
 ``The Swampland Spectrum Conjecture in Inflation,''
 arXiv:1810.11699 [hep-th].
 
 \bibitem{quintessence}
 R.~R.~Caldwell, R.~Dave and P.~J.~Steinhardt,
 ``Cosmological imprint of an energy component with general equation of state,''
 Phys.\ Rev.\ Lett.\  {\bf 80}, 1582 (1998)
 doi:10.1103/PhysRevLett.80.1582
 [astro-ph/9708069].
 
 \bibitem{Lavinia}
 L.~Heisenberg, M.~Bartelmann, R.~Brandenberger and A.~Refregier,
 ``Dark Energy in the Swampland,''
 Phys.\ Rev.\ D {\bf 98}, no. 12, 123502 (2018)
 doi:10.1103/PhysRevD.98.123502
 [arXiv:1808.02877 [astro-ph.CO]];\\
 L.~Heisenberg, M.~Bartelmann, R.~Brandenberger and A.~Refregier,
 ``Dark Energy in the Swampland II,''
 arXiv:1809.00154 [astro-ph.CO].
 
 \bibitem{Linde}
 Y.~Akrami, R.~Kallosh, A.~Linde and V.~Vardanyan,
 ``The landscape, the swampland and the era of precision cosmology,''
 Fortsch.\ Phys.\  {\bf 2018}, 1800075
 doi:10.1002/prop.201800075
 [arXiv:1808.09440 [hep-th]];\\
 M.~C.~David Marsh,
 ``The Swampland, Quintessence and the Vacuum Energy,''
 Phys.\ Lett.\ B {\bf 789}, 639 (2019)
 doi:10.1016/j.physletb.2018.11.001
 [arXiv:1809.00726 [hep-th]];\\
 H.~Fukuda, R.~Saito, S.~Shirai and M.~Yamazaki,
 ``Phenomenological Consequences of the Refined Swampland Conjecture,''
 arXiv:1810.06532 [hep-th];\\
 S.~K.~Garg, C.~Krishnan and M.~Zaid,
 ``Bounds on Slow Roll at the Boundary of the Landscape,''
 arXiv:1810.09406 [hep-th];\\
 P.~Agrawal and G.~Obied,
 ``Dark Energy and the Refined de Sitter Conjecture,''
 arXiv:1811.00554 [hep-ph];\\
 C.~I.~Chiang, J.~M.~Leedom and H.~Murayama,
 ``What does Inflation say about Dark Energy given the Swampland Conjectures?,''
 arXiv:1811.01987 [hep-th].
 
 \bibitem{Raveri}
 M.~Raveri, W.~Hu and S.~Sethi,
 ``Swampland Conjectures and Late-Time Cosmology,''
 arXiv:1812.10448 [hep-th].
 
 
 \bibitem{Heisenberg:2019qxz}
 L.~Heisenberg, M.~Bartelmann, R.~Brandenberger and A.~Refregier,
 ``Horndeski in the Swampland,''
 arXiv:1902.03939 [hep-th].
 
 
 \bibitem{stoch}
 A.~A.~Starobinsky,
 ``Stochastic De Sitter (inflationary) Stage In The Early Universe,''
 Lect.\ Notes Phys.\  {\bf 246}, 107 (1986).
 doi:10.1007/3-540-16452-96
 
 \bibitem{eternal}
 A.~D.~Linde,
 ``Eternal Chaotic Inflation,''
 Mod.\ Phys.\ Lett.\ A {\bf 1}, 81 (1986);
 doi:10.1142/S0217732386000129\\
 M.~Aryal and A.~Vilenkin,
 ``The Fractal Dimension of Inflationary Universe,''
 Phys.\ Lett.\ B {\bf 199}, 351 (1987).
 doi:10.1016/0370-2693(87)90932-4
 
 \bibitem{large}
 H.~Matsui and F.~Takahashi,
 ``Eternal Inflation and Swampland Conjectures,''
 Phys.\ Rev.\ D {\bf 99}, no. 2, 023533 (2019)
 doi:10.1103/PhysRevD.99.023533
 [arXiv:1807.11938 [hep-th]];\\
 K.~Dimopoulos,
 ``Steep Eternal Inflation and the Swampland,''
 Phys.\ Rev.\ D {\bf 98}, no. 12, 123516 (2018)
 doi:10.1103/PhysRevD.98.123516
 [arXiv:1810.03438 [gr-qc]].
 
 \bibitem{Kinney}
 W.~H.~Kinney,
 ``Eternal Inflation and the Refined Swampland Conjecture,''
 arXiv:1811.11698 [astro-ph.CO].
 
 \bibitem{Bousso}
 R.~Bousso, B.~Freivogel and I.~S.~Yang,
 ``Eternal Inflation: The Inside Story,''
 Phys.\ Rev.\ D {\bf 74}, 103516 (2006)
 doi:10.1103/PhysRevD.74.103516
 [hep-th/0606114].
 
 \bibitem{Shandera}
 S.~Brahma and S.~Shandera,
  ``Stochastic eternal inflation is in the swampland,''
  arXiv:1904.10979 [hep-th].
  
  \bibitem{Rudelius}
  T.~Rudelius,
  ``Conditions for (No) Eternal Inflation,''
  arXiv:1905.05198 [hep-th].
  
 \bibitem{swamp1}
 H.~Ooguri and C.~Vafa,
 ``On the Geometry of the String Landscape and the Swampland,''
 Nucl.\ Phys.\ B {\bf 766}, 21 (2007)
 doi:10.1016/j.nuclphysb.2006.10.033
 [hep-th/0605264].
 
 \bibitem{swamp2}
 G.~Obied, H.~Ooguri, L.~Spodyneiko and C.~Vafa,
 ``De Sitter Space and the Swampland,''
 arXiv:1806.08362 [hep-th].
 
 \bibitem{Berera}
 A.~Berera,
 ``Warm inflation,''
 Phys.\ Rev.\ Lett.\  {\bf 75}, 3218 (1995)
 doi:10.1103/PhysRevLett.75.3218
 [astro-ph/9509049].
 
 \bibitem{warm}
 S.~Das,
 ``Warm Inflation in the light of Swampland Criteria,''
 arXiv:1810.05038 [hep-th];\\
 M.~Motaharfar, V.~Kamali and R.~O.~Ramos,
 ``Warm way out of the Swampland,''
 arXiv:1810.02816 [astro-ph.CO].
 
 \bibitem{swamp4}
 H.~Ooguri, E.~Palti, G.~Shiu and C.~Vafa,
 ``Distance and de Sitter Conjectures on the Swampland,''
 arXiv:1810.05506 [hep-th];
 
 \bibitem{GH}
 G.~W.~Gibbons and S.~W.~Hawking,
 ``Cosmological Event Horizons, Thermodynamics, and Particle Creation,''
 Phys.\ Rev.\ D {\bf 15}, 2738 (1977).
 doi:10.1103/PhysRevD.15.2738
 
 \bibitem{other2}
 S.~K.~Garg and C.~Krishnan,
 ``Bounds on Slow Roll and the de Sitter Swampland,''
 arXiv:1807.05193 [hep-th];\\
 F.~Denef, A.~Hebecker and T.~Wrase,
 ``The dS swampland conjecture and the Higgs potential,''
 arXiv:1807.06581 [hep-th];\\
 D.~Andriot,
 Phys.\ Lett.\ B {\bf 785}, 570 (2018)
 doi:10.1016/j.physletb.2018.09.022
 [arXiv:1806.10999 [hep-th]];\\
 D.~Andriot,
 ``New constraints on classical de Sitter: flirting with the swampland,''
 Fortsch.\ Phys.\  {\bf 67}, no. 1-2, 1800103 (2019)
 doi:10.1002/prop.201800103
 [arXiv:1807.09698 [hep-th]];\\
 C.~Roupec and T.~Wrase,
 ``de Sitter extrema and the swampland,''
 arXiv:1807.09538 [hep-th];\\
 J.~P.~Conlon,
 ``The de Sitter swampland conjecture and supersymmetric AdS vacua,''
 Int.\ J.\ Mod.\ Phys.\ A {\bf 33}, no. 29, 1850178 (2018)
 doi:10.1142/S0217751X18501786
 [arXiv:1808.05040 [hep-th]];\\
 K.~Dasgupta, M.~Emelin, E.~McDonough and R.~Tatar,
 ``Quantum Corrections and the de Sitter Swampland Conjecture,''
 JHEP {\bf 1901}, 145 (2019)
 doi:10.1007/JHEP01(2019)145
 [arXiv:1808.07498 [hep-th]];\\
 U.~Danielsson,
 ``The quantum swampland,''
 arXiv:1809.04512 [hep-th];\\
 D.~Andriot and C.~Roupec,
 ``Further refining the de Sitter swampland conjecture,''
 Fortsch.\ Phys.\  {\bf 67}, no. 1-2, 1800105 (2019)
 doi:10.1002/prop.201800105
 [arXiv:1811.08889 [hep-th]];\\
 K.~Hamaguchi, M.~Ibe and T.~Moroi,
 ``The swampland conjecture and the Higgs expectation value,''
 JHEP {\bf 1812}, 023 (2018)
 doi:10.1007/JHEP12(2018)023
 [arXiv:1810.02095 [hep-th]];\\
 A.~Hebecker and T.~Wrase,
 ``The asymptotic dS Swampland Conjecture - a simplified derivation and a potential loophole,''
 Fortsch.\ Phys.\  {\bf 2018}, 1800097
 doi:10.1002/prop.201800097
 [arXiv:1810.08182 [hep-th]];\\
 A.~Banlaki, A.~Chowdhury, C.~Roupec and T.~Wrase,
 ``Scaling limits of dS vacua and the swampland,''
 arXiv:1811.07880 [hep-th];\\
 D.~Junghans,
 ``Weakly Coupled de Sitter Vacua with Fluxes and the Swampland,''
 arXiv:1811.06990 [hep-th];\\
 P.~Corvilain, T.~W.~Grimm and I.~Valenzuela,
 ``The Swampland Distance Conjecture for Kahler moduli,''
 arXiv:1812.07548 [hep-th].
 
 \bibitem{other3}
 A.~Kehagias and A.~Riotto,
 ``A note on Inflation and the Swampland,''
 arXiv:1807.05445 [hep-th];\\
 H.~Murayama, M.~Yamazaki and T.~T.~Yanagida,
 ``Do We Live in the Swampland?,''
 JHEP {\bf 1812}, 032 (2018)
 doi:10.1007/JHEP12(2018)032
 [arXiv:1809.00478 [hep-th]];\\
 M.~Ibe, M.~Yamazaki and T.~T.~Yanagida,
 ``Quintessence Axion from Swampland Conjectures,''
 arXiv:1811.04664 [hep-th];\\
 J.~J.~Blanco-Pillado, M.~A.~Urkiola and J.~M.~Wachter,
 ``Racetrack Potentials and the de Sitter Swampland Conjectures,''
 JHEP {\bf 1901}, 187 (2019)
 doi:10.1007/JHEP01(2019)187
 [arXiv:1811.05463 [hep-th]];\\
 M.~Emelin and R.~Tatar,
 ``Axion Hilltops, Kahler Modulus Quintessence and the Swampland Criteria,''
 arXiv:1811.07378 [hep-th];\\
 C.~M.~Lin,
 ``Type I Hilltop Inflation and the Refined Swampland Criteria,''
 Phys.\ Rev.\ D {\bf 99}, no. 2, 023519 (2019)
 doi:10.1103/PhysRevD.99.023519
 [arXiv:1810.11992 [astro-ph.CO]];\\
 S.~C.~Park,
 ``Minimal gauge inflation and the refined Swampland conjecture,''
 JCAP {\bf 1901}, no. 01, 053 (2019)
 doi:10.1088/1475-7516/2019/01/053
 [arXiv:1810.11279 [hep-ph]];\\
 D.~Y.~Cheong, S.~M.~Lee and S.~C.~Park,
 ``Higgs Inflation and the Refined dS Conjecture,''
 Phys.\ Lett.\ B {\bf 789}, 336 (2019)
 doi:10.1016/j.physletb.2018.12.046
 [arXiv:1811.03622 [hep-ph]];\\
 M.~Scalisi and I.~Valenzuela,
 ``Swampland Distance Conjecture, Inflation and $\alpha$-attractors,''
 arXiv:1812.07558 [hep-th].
 
 \bibitem{Dvali}
 G.~Dvali and C.~Gomez,
 ``Quantum Exclusion of Positive Cosmological Constant?,''
 Annalen Phys.\  {\bf 528}, 68 (2016)
 doi:10.1002/andp.201500216
 [arXiv:1412.8077 [hep-th]];\\
 G.~Dvali and C.~Gomez,
 ``On Exclusion of Positive Cosmological Constant,''
 Fortsch.\ Phys.\  {\bf 67}, no. 1-2, 1800092 (2019)
 doi:10.1002/prop.201800092
 [arXiv:1806.10877 [hep-th]];\\
 G.~Dvali, C.~Gomez and S.~Zell,
 ``Quantum Breaking Bound on de Sitter and Swampland,''
 Fortsch.\ Phys.\  {\bf 67}, no. 1-2, 1800094 (2019)
 doi:10.1002/prop.201800094
 [arXiv:1810.11002 [hep-th]].
 
 \bibitem{WGC}
 N.~Arkani-Hamed, L.~Motl, A.~Nicolis and C.~Vafa,
 ``The String landscape, black holes and gravity as the weakest force,''
 JHEP {\bf 0706}, 060 (2007)
 doi:10.1088/1126-6708/2007/06/060
 [hep-th/0601001].
 
 \bibitem{SGC}
 R.~H.~Brandenberger,
 ``String Gas Cosmology,''
 String Cosmology, J.Erdmenger (Editor).  Wiley, 2009. p.193-230
 [arXiv:0808.0746 [hep-th]].
 
 \bibitem{BNPV}
 R.~H.~Brandenberger, A.~Nayeri, S.~P.~Patil and C.~Vafa,
 ``Tensor Modes from a Primordial Hagedorn Phase of String Cosmology,''
 Phys.\ Rev.\ Lett.\  {\bf 98}, 231302 (2007)
 doi:10.1103/PhysRevLett.98.231302
 [hep-th/0604126].
 
 \bibitem{Patil}
 S.~P.~Patil and R.~H.~Brandenberger,
 ``The Cosmology of massless string modes,''
 JCAP {\bf 0601}, 005 (2006)
 doi:10.1088/1475-7516/2006/01/005
 [hep-th/0502069];\\
 S.~P.~Patil and R.~Brandenberger,
 ``Radion stabilization by stringy effects in general relativity,''
 Phys.\ Rev.\ D {\bf 71}, 103522 (2005)
 doi:10.1103/PhysRevD.71.103522
 [hep-th/0401037];\\
 S.~Watson and R.~Brandenberger,
 ``Stabilization of extra dimensions at tree level,''
 JCAP {\bf 0311}, 008 (2003)
 doi:10.1088/1475-7516/2003/11/008
 [hep-th/0307044].
 
 \bibitem{MFB}
 V.F. Mukhanov, H.A. Feldman and R.H. Brandenberger,
 ``Theory of Cosmological Perturbations''
 Physics Reports \textbf{215}, 203 (1992).
 
 \bibitem{RHBrev}
 R.~H.~Brandenberger,
 ``Lectures on the theory of cosmological perturbations,''
 Lect.\ Notes Phys.\  {\bf 646}, 127 (2004)
 doi:10.1007/978-3-540-40918-25
 [hep-th/0306071].
 
 \bibitem{Guth2}
 A.~H.~Guth and E.~J.~Weinberg,
 ``Could the Universe Have Recovered from a Slow First Order Phase Transition?,''
 Nucl.\ Phys.\ B {\bf 212}, 321 (1983).
 doi:10.1016/0550-3213(83)90307-3
 
\end{thebibliography}
\end{document}